      [1999/12/01 v1.4c Il Nuovo Cimento]
\def\ltsima{$\; \buildrel < \over \sim \;$}
\def\gtsima{$\; \buildrel > \over \sim \;$}
\def\simlt{\lower.5ex\hbox{\ltsima}}
\def\simgt{\lower.5ex\hbox{\gtsima}}
\def\pmb#1{\setbox0=\hbox{#1}%
    \kern-.025em\copy0\kern-\wd0
    \kern.05em\copy0\kern-\wd0
    \kern-.025em\raise.0433em\box0 }

\documentclass{cimento}

\usepackage{graphicx}  
\title{CMB Anisotropies: Their Discovery and Utilization}
\author{George~F. Smoot}
\instlist{\inst{ }Lawrence Berkeley National Laboratory and Berkeley Center for Cosmological Physics 

Physics Department, 
University of California, Berkeley CA 94720}
\begin{document}

\maketitle

\begin{abstract}
  This article is a written and modified version of a talk presented at
  the conference `{\it A Century of Cosmology}' held at San Servolo, Venice,
  Italy, in August 2007.  
  The talk focuses on some of the cosmology history leading to the discovery and exploitation 
  of Cosmic Microwave Background (CMB) Radiation anisotropies.
 We have made tremendous advances first in the development of the techniques to 
 observe these anisotropies and in observing and interpreting them to extract their contained cosmological information.
 CMB anisotropies are now a cornerstone in our understanding of the cosmos and our future progress in the field.
 This is an outcome that Dennis Sciama hoped for and encouraged.
\end{abstract}

\begin{figure}
\centering
\includegraphics[width=1.1\textwidth]{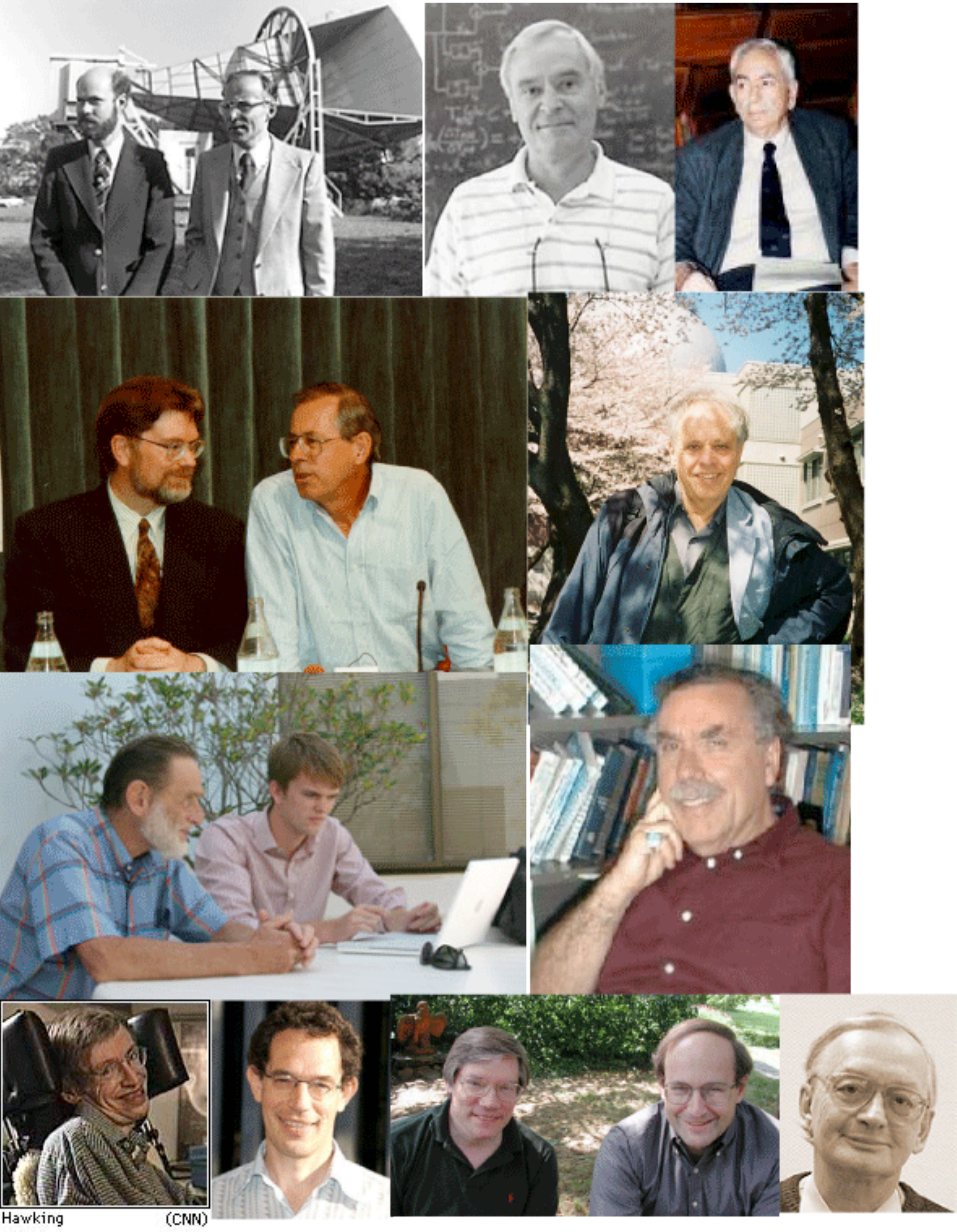}    
\vskip -0.1in
\caption{Photos of  Robert Wilson  and Arno Penzias in front of the Bell Labs Homdel antenna used for their observations and discovery of the CMB,  David Wilkinson, and Dennis Sciama.
Second row shows author George Smoot with Jim Peebles and Joe Silk.
Third row shows Rainer Sachs with student Chris Reed and Art Wolfe. 
Fourth and bottom row shows Stephen Hawking, Neil Turok,  Alan Guth and Paul Steinhardt, Alexi Starobinskii. }
\end{figure}

\section{Introduction}
It was a privilege to be invited to give a talk at this San Servolo 
meeting celebrating {`\it A Century of Cosmology'} and commemorating Dennis Sciama. 

Perhaps the most conclusive (and certainly among the most carefully examined) piece of evidence for Big Bang Cosmology is the existence of an isotropic radiation bath that permeates the entire Universe,  known as the``cosmic microwave background" (CMB). 
The word ``isotropic" means the same in all directions.
In 1964, two young radio astronomers, Arno Penzias and Robert Wilson, working at Bell Laboratories discovered the CMB using a  well-calibrated horn antenna. 
They soon determined that the radiation was diffuse, emanated uniformly from all directions in the sky, and had a temperature of approximately 3 Kelvin (i.e. 3$^\circ$C above absolute zero). 
Initially, they could find no satisfactory explanation for their observations, and considered the possibility that their signal may have been due to some undetermined systematic effect or noise. 
Their careful work soon eliminated alternative possibilities.
Then MIT radio astronomer Bernie Burke drew their attention to Robert Dicke and Jim Peebles of Princeton who were considering the existence of a potential relic radiation and they had encouraged colleagues Peter Roll and David  Wilkinson to begin constructing an instrument and start a program to search for this relic radiation. 
Eventually they learned that this background radiation had in fact been predicted years earlier by George Gamow\cite{ref:Gamow48} and colleagues Ralph Alpher and Robert Hermann as a relic of the evolution of the early Universe. 
This background of microwaves was the cooled remnant of the primeval fireball - an echo of the Big Bang.

\begin{figure}
\centering
\end{figure}

We have learned with increasing precision since 1964 that, on the largest scales, 
the Universe is highly uniform. 
This understanding began with 1964, Arno Penzias and Robert Wilson of Bell Laboratories in New Jersey discovering\cite{ref:Penzias65} the cosmic background radiation and the determination that it was isotropic to at least the 10\%\ level.  
This radiation is a relic from the hot big bang, which cooled to its present very low temperature of 2.7 Kelvin as the Universe expanded. 
Within a year of this discovery for which  Penzias and Wilson were awarded the Nobel Prize for Physics in 1978, experiments showed that the temperature of the cosmic background is the same in every direction to within a few per cent.  
Much of this early work consisted of follow up experiments involving David Wilkinson.
While I was a graduate student, David Wilkinson\cite{ref:Wilkinson67} at Princeton 
began a program to observe the isotropy and the anisotropy of the CMB and 
started training the first of multiple generations of students.
At the same time theorists Joe Silk\cite{ref:Silk67}, Jim Peebles\cite{ref:Peebles67}, Rainer Sachs and Art Wolfe\cite{ref:Sachs67}  and others were providing the beginnings of understanding and motivation for these observations.

So, unless we live in a spherically symmetric universe centered with exquisite precision on the Earth - which appears unlikely -  we can conclude that the Universe must be very nearly uniform on large scales. 
The problem of how to reconcile the large-scale uniformity of the Universe with its apparent clumpy structure on small scales has baffled cosmologists for decades. 

In 1967 Joe Silk pointed out the need for primordial fluctuations - the seeds of this structure -  and the difficulty of their surviving the primeval fireball that must have existed given the 3 K relic radiation's cosmological origin. 
It was Silk's 1967 description that first drew my attention to the important and dominant role that the CMB had in the early universe in addition to being a present day relic. 
I also found additional motivation in papers by Sciama and by Hawking about what the CMB could reveal to us about the largest scale structure of the Universe.

There is no generally accepted picture for the formation of structure in the Universe, though 
most cosmologists now subscribe to the idea that there was an early period of accelerating expansion at very high speeds and quantum mechanical microscopic fluctuations were stretched by the tremendous expansion of the universe to cosmological scales including to sizes of the galaxies and clusters of galaxies.
We call this first portion Inflation and the second step is gravitational attraction or instability to form structure.
Over-densities in the Universe gravitationally attracted the material around them, 
so small irregularities were amplified by gathering matter and energy.
Eventually, these regions of over densities became large enough and dense enough to collapse and to form galaxies, which themselves clumped under the influence of gravity to form clusters and superclusters of galaxies, and so on. 
This idea, that gravity gave the Universe its structure, 
is almost as old as Newton's law of gravitation itself.

A description of how such gravitational clumping works in an expanding universe was first given by the Soviet physicist Evgenii Lifshitz in 1946. 
He found that clumps amplified more slowly in an expanding universe because gravity has to fight against the universal expansion. 
Fluctuations grow linearly with the expansion of the universe rather than exponentially.
As a result, rather large fluctuations in density were needed to make the structure that we see today. 
At the time, physicists had no idea where such fluctuations could have come from. 
Although cosmologists have suggested other ideas, for example that galaxies formed from some kind of `cosmic turbulence', gravity seems the most plausible explanation for the growth of structure in the universe.

The most compelling case in favor of gravity was made by Jim Peebles of Princeton University. 
In the 1970s Peebles and his collaborators discovered statistical evidence that galaxies cluster in a heirarchical way, and showed how this order could be explained by gravitational clustering. 
According to Peebles, given the right kind of irregularities in the early Universe, gravity could explain both the observed structure, and the order in which it has formed (galaxies before clusters before superclusters). 
The next step was to develop a theory to explain how the right kind of irregularities might arise, and to design experiments to measure them on the largest possible scales.

A few years earlier, shortly after the discovery of the cosmic background radiation, Sachs and Wolfe\cite{ref:Sachs67},  Silk (then at Harvard)\cite{ref:Silk67}, Peebles, the Soviet physicist Yakov Zeldovich and others had realised that irregularities in the early Universe would lead to tiny, but perhaps measurable, temperature differences in the radiation coming from different directions in the sky. 
There are several physical processes that can cause such ripples, but the simplest and most important is the effect described by Sachs and Wolfe, which bears their name. 
This is that light moving in a gravitational field can experience small wavelength changes (redshifts and blueshifts).

This effect, which is predicted by the theory of General Relativity, is that there will be a frequency change proportional to the change in gravitational potential divided by the speed of light squared.
It has been observed in the light from stars and in experiments on Earth. 
Because of irregularities in the early Universe, some of the photons of the background radiation will have experienced tiny net redshifts or blueshifts on their way to Earth since they travel out of different potential wells though as Sachs and Wolfe had shown there were more photons beginning in the potential wells so that the effect is less by a factor of three. 
By measuring these Sachs-Wolfe fluctuations it is possible to map irregularities as they were when the Universe became transparent to photons - that is, when it was only about a four tenths of a million years old. 
The search for these temperature fluctuations became a prime goal, or as Michael Turner of the University of Chicago put it, the `Holy Grail' of cosmology.

Before COBE, the only temperature variation observed was the `dipole anisotropy', a discrepancy in the temperature arising from the motion of our Galaxy through the sea of cosmic background photons. 
This was discovered in 1977 by the author George Smoot and colleagues from Berkeley using sensitive radiometers mounted on a U2 spy plane. 
But apart from this, by the 1980s observations, particularly from our Berkeley and Wilkinson's group balloon-borne experiments, showed that the background radiation temperature was embarrassingly smooth. 
There was apparently not even a temperature variation of around 0.01 per cent, especially on the largest angular scales.  
If, as Lifshitz had said, density fluctuations of about a tenth of a per cent were needed in the early Universe to make galaxies by the present day, 
why was there no trace of these fluctuations in the cosmic background?

\subsection{CDM - COLD DARK MATTER}
in 1933, Fritz Zwicky of the California Institute of Technology began arguing that large clusters of galaxies could not be held together by gravity unless most of their mass was in an unknown `dark' form. Later on cosmologists speculated that such dark matter might be made up of unusual particles - not the protons, neutrons and electrons of ordinary matter, but more exotic `weakly interacting' particles such as the neutrino, which was detected in 1956. In 1980, several cosmologists realised that exotic dark matter could explain the lack of temperature ripples in the cosmic background radiation.

They argued that whereas electrons and protons are electromagnetically coupled to the cosmic background radiation, weakly interacting particles are not. 
They feel only the weak force responsible for radioactive decay and the weaker force of gravity, so they can begin to cluster under the action of gravity earlier than ordinary matter which was held apart by the cosmic background  radiation. 
The clustering of the dark matter could begin gravitational clustering about ten times earlier than the baryons and thus could arise from ten times smaller density perturbations.
Thus in a universe dominated by weakly interacting particles, smaller initial irregularities are needed to make the structure that we see today and the ripples in the background radiation would be too small to have been detected up to that point.
But theoretical models based on low-mass neutrinos could not be made to work. 
Neutrinos with a small mass would be `hot' in the early Universe, moving at close to the speed of light. Because of these motions, small irregularities in the distribution of matter would be washed out, leaving only structures more massive than large clusters of galaxies, in conflict with observations. 
However, cosmologists soon realised that the dark matter could be made of even more exotic things: supersymmetric particles such as the gravitino or photino, for example, or oscillating scalar fields such as the axion. 
Unlike neutrinos, these could be `cold' in the early Universe, giving rise to structure over a wide range of scales, from dwarf galaxies to the giant superclusters and voids. 
This theory of `cold dark matter' (CDM) proved very successful. 
At last there was a theory which could explain a wide range of phenomena and which offered a framework for detailed theoretical calculations. 
Among other things, it meant theoreticians could calculate the expected pattern of fluctuations in the background radiation quite precisely. 

At the same time measurements of the velocities of galaxies indicated that the potential wells would be at the level needed for CDM to be the correct model (c.f. Gorski\cite{ref:Gorski1991}) and at the levels just below that accessible by the COBE DMR instrument. 
Due to the long delay in launching COBE, as PI, I had realized that both the theoretical requirements and developments in technology made it both necessary and possible to improve the DMR instrument, if we were to reach our goals successfully. 
The original DMR was to measure temperature differences between regions of the sky 60$^\circ$ apart at four frequencies: 23.5, 31.5, 53 and 90 Gigahertz (wavelengths of 12.8, 9.5, 5.7 and 3.3 millimetres). 
This need was pitched first to the team, then the larger COBE science team, and then to the NASA GSFC and Headquarters management. 
The improved DMR measured temperature differences between regions of the sky 60$^\circ$ apart at three frequencies: 31.5, 53 and 90 Gigahertz 
with the 53 and 90 GHz passively cooled to about 135 K which resulted in more than a factor of two less noise and thus a factor of two improved sensitivity.

In  April 1992 our COBE DMR  team made our announcement\cite{ref:Smoot92}. 
The analysis of the first year's data from the differential microwave radiometer (DMR), one of three experiments on board COBE, showed evidence of the long sought CMB fluctuations.
From the first year's data, we were able to make maps of the entire sky at each of the three frequencies sampled by the DMR. 
The maps showed again that the sky is incredibly uniform but there were some features.
The most conspicuous features are the dipole anisotropy and emission from our own Galaxy, clearly visible as a thin band (the galactic plane) running through the middle of each map. 
When the best-fitted dipole pattern is subtracted, the maps show `ripples' - residual temperature variations of 0.0011 per cent (30.5 microkelvin) - spread around the sky in regions away from the Galaxy.

By comparing the three maps, we were also able to show, crucially, that the temperature fluctuations were independent of the frequency of the radiation. 
This must be true on physical grounds, if they were indeed caused by the `seed' fluctuations that made the observed structure in the Universe. 
We know from the COBE maps and from other surveys that the Galaxy emits radiation at millimeter wavelengths. 
But it is difficult to predict exactly how much emission there would be high above the galactic plane. 
In principle, the temperature fluctuations in the long wavelength map could have been caused by electromagnetic radiation produced by cosmic ray particles spiralling around in the Galaxy's magnetic field. 
At short wavelengths the fluctuations might have been caused by emission from dust in the Galaxy. 
A chance agreement of all three maps could not be ruled out entirely 
until the fluctuations were detected by other experiments covering a wider range of wavelengths and angular scales. 

By assuming that the fluctuations were primordial, what did we learn? 
The CMB fluctuations observed by COBE have angular sizes on the sky greater than 10$^\circ$. 
This corresponds to enormous structures at early times, which must measure more than 3000 million light years across today. 
The existence of such structures in the early Universe contradicts our usual notions of cause and effect. 
The ripples are so large that by the time the photons detected by COBE were emitted, the Universe was simply not old enough for a light signal to have crossed from one side of a COBE-observed fluctuation to the other. 
What physical process could have led to such large lumps so early?

This may be the most significant aspect of the COBE discovery. 
It points to new phenomena, well beyond those described by the standard Glashow-Weinberg-Salam model of particle physics, and provides vital clues to the mechanism that created the fluctuations in the extremely young Universe. 
During the 1980s, two theories emerged to account for these fluctuations. 
Both are based on highly speculative physics that might apply at energies more than 12 orders of magnitude higher than the highest energies achieved in particle accelerators on Earth.

\subsection{INFLATION}
In 1980 Alan Guth, now of the Massachusetts Institute of Technology, proposed the idea of an `inflationary phase' - a period of extremely rapid expansion to solve issues such as the observed absence of magnetic monopoles, the flatness of the universe and the large scale isotropy of the universe.
A short time later papers by Steinhardt \&\ Albrecht\cite{ref: Albrecht} and Linde\cite{ref:Linde} proposed models with simple scalar fields that showed the idea of Inflation could be made to work in principle.

According to Inflation theory, the seeds which gave rise to the present day structure in the Universe are quantum fluctuations,  minute variations on a subatomic scale,  that were expanded by more than 60 orders of magnitude as the Universe went through the period of inflation and subsequent expansion. 
At the 1982 Nuffield conference Guth, Alexei Starobinsky, Stephen Hawking, Paul Steinhardt, Mike Turner, James Bardeen and others showed most models of inflation predict that these seed fluctuations are generated during inflation and would have a characteristic `scale invariant' form\cite{ref:Bardeen}. 
That is, the amplitude of the fluctuations in the gravitational field is independent of their physical size. And these fluctuations would, by the Sachs-Wolfe effect mentioned earlier, produce a characteristic `scale invariant' pattern of temperature fluctuations on the sky, which fits in with what COBE saw.

\subsection{TOPOLOGICAL DEFECTS}

According to the second theory, the seed fluctuations arose during a `phase transition' in which the Universe passed from one high energy phase to another at lower energy. 
This is similar to the phase transition from water to ice as it cools. 
In the early Universe, regions of the high energy phase can become trapped during the phase transition, as a result of a symmetry breaking. 
These regions form `topological defects' in the fabric of space, some of which might stretch right across the observable Universe. 

These topological defects include exotic structures called domain walls, cosmic strings and cosmic textures. 
The particular type of defect depends on what symmetry was broken in the phase transition. 
Domain walls are two-dimensional (sheet-like) objects, but their existence is ruled out in the early Universe because they would have produced unacceptably large fluctuations at the present day. Cosmic strings are line-like objects with a mass per unit length of as much as 10$^{17}$ tons per meter and a thickness of only 10$^{- 22}$ of the radius of a hydrogen atom if they came from the GUT scale. 
At first it was thought that galaxies and clusters would form around cosmic strings.
Computer simulations of the evolution of a string network show that strings cut each other up into loops that decay away by radiating gravitational waves. 
This lets them scale so as not to come to dominate the energy in the Universe.
 It was possible that large-scale structure forms in the wakes produced behind long strings. 

Neil Turok of Princeton University and Imperial College, London proposed in 1989 cosmic textures as a possible alternative. 
Textures are a more complicated type of defect that can arise in some unified theories of particle physics - a type of higher dimensional topological knot that unwinds at the speed of light. 
As the texture shrinks, it attracts matter towards it, producing density perturbations that are then amplified by gravity. 
As Turok and his colleagues showed at the time, a texture-seeded universe dominated by cold dark matter could match the features observed to that point in Galaxy surveys.

\subsection{DATA CONFRONTS THEORY}
Surprisingly, these very different models for the origin of irregularities in the early Universe predict almost identical `scale invariant' fluctuations in the cosmic background radiation on large angular scales.  
According to theory, the fluctuations predicted in the texture model should show slightly greater contrast than those of the inflationary model.
The texture-induced fluctuations are non-Gaussian to a significant degree. 
But at the low resolution of the COBE DMR experiment, these differences are almost imperceptible. Instead, we had to ask whether the amplitudes of the temperature differences measured by COBE could distinguish between these models. 
The COBE signal is about twice what we would have expected in the simplest CDM universe with inflationary perturbations. 
The difference was thought to be greater in a texture-seeded CDM model. 
Neither theory seemed to fit the the observations exactly but simple CDM seemed best.

By a year or two after the discovery several of us noticed that the COBE DMR observations were a factor of two of the minimum, but by that time also the maximum predicted by the CDM model. 
We should have then followed up on that but two things stopped us: (1) first the great relief that we had finally found the fluctuations rather than a limit below the minimum required for gravity and (2) the fact that the supporting observations were not yet sufficient to really define things starkly.

We needed more and better observations of galaxy distributions as pioneered by Peebles.
The Center for Astrophysics Redshift Survey was started in 1977 by Marc Davis, John Huchra, Dave Latham and John Tonry. The First CfA Survey, completed in 1982\cite{ref:Huchra83}. 
These observations showed surprising large scale features but were insufficient in coverage.
This motivated the next generation of surveys on a larger scale.

During the 1980s a number of groups, for example George Efstathiou's  at Oxford and others around Great Britain composing 2dFGRS, were measuring large-scale structure in the galaxy distribution, using ground based telescopes. 
These projects were designed to detect lumps that are 100 million light years or more across. 
This is the transition area where structure becomes almost imperceptible above the mean background density of the Universe. 
These surveys were giving increasingly strong evidence that the Universe is more lumpy than the simplest CDM models predict. 
Observations showing that galaxies are moving rapidly on large scales have presented additional evidence for large clumps of dark matter in the Universe.

At least three variants of the CDM theory were proposed to explain these results. 
Each involves tinkering with a key ingredient of the standard model. 
For example, it is easy to invent models of inflation that produce more large-scale structure than in the standard model, though none of these seem particularly compelling, because all involve a coincidence of some sort. 
Another possibility is that the cosmological constant, which Einstein added to his equations to produce a static Universe, is non-zero.

\subsection{$\Lambda$CDM}
In 1993 it was recognized that both the COBE ripples and the large-scale clustering of galaxies can be explained by a $\Lambda$CDM universe in which 80 per cent of the present mass density is contributed by a cosmological constant, though some cosmologists argued that such theories may not explain the motions of galaxies.
A large section of cosmologists thought this too ugly a model to possibly be correct and could not understand the motivation for a non-zero cosmological constant.
As a result this model was not to be pursued actively until half a decade later.

\subsection{DARK BLEND -- HCDM}

The theory that first emerged as the favorite involved a `cocktail' of dark matter. 
A Universe that contains about 60 per cent cold dark matter, 30 per cent hot dark matter, assumed to be in the form of low-mass 'tau' neutrinos, and 10 per cent in ordinary (baryonic) particles like protons and neutrons, seems to fit all the observations. 
More accurate measurements of matter distribution on large scales were needed to test these modified CDM theories.

Many cosmologists had the view that one should not be too disturbed about small discrepancies between theory and observation. 
Any attempt to explain cosmic structure involves extrapolating from the Planck era, 10$^{- 43}$ seconds after the big bang, when effects like quantum gravity were important, to the present Universe which is about 14 billion years old. 
So it would not surprising that one or two things do not seem to fit. 
Theorists knew too little about the early Universe and the nature of dark matter to make certain predictions. 
The COBE observations were important because they provided theorists with a precious link to the early Universe. 
The CMB fluctuations allow us to study physics at ultra-high energies that may eventually explain why our Universe is as it is.

\subsection{CMB ANISOTROPY OBSERVATIONS IN HIGH GEAR }
With the COBE DMR discovery and the theoretical promise of observations for telling us about the early universe, those in the field were joined by many new talented people and all were highly motivated to continue observations.
The expectation was that such observations should shed light on some of cosmology's most pressing problems. 
Was there a period of inflation in the early Universe? 
Were seed irregularities generated from quantum fluctuations or from topological defects? 
What is the nature of the dark matter? 
Enormous projects were also planned to map the distribution of galaxies on large scales. 

Within the next few years  after the COBE DMR announcement, the primordial ripples in the background radiation were observed by ground-based and balloon-borne experiments over a wide range of angular scales. 
Only a few months after COBE DMR a balloon-borne experiment, FIRS, by Stephen Meyer of MIT, Ed Cheng of the Goddard Space Flight Center in Greenbelt, Maryland and Lyman Page of Princeton University saw similar ripples in the same region of the sky to those seen by COBE. 
This experiment detects radiation of millimeter and submillimeter wavelengths, a different range from COBE, providing further evidence that the ripples are primordial, and are not just background emission from our own Galaxy.

Another experiment carried out in 1990 at the South Pole by Phil Lubin of the University of California at Santa Barbara and his colleagues first had a lower limit and was close to disagreeing with COBE but with more data analysis they decided that it may have detected primordial ripples at angular scales of 1.5 degrees, though they were not at first convinced that they could rule out a galactic origin for their signal. 
COBE continued to make observations, and the analysis of the second year's data, and full four years' data ultimately confirmed and improved the original DMR results. 
Other experiments, such as the one at Tenerife by Rod Davies of Jodrell Bank and his collaborators, a team that includes scientists from Cambridge and the Instituto Astrofisica de Canarias found matching detecting primordial ripples to COBE DMR's.

The COBE discovery results stood up well to examination and  received support from other experiments. 
It did seem as though the CMB fluctuations were real, and that a new era in cosmology had begun.

\subsection{A ROUND WORLD IN A FLAT UNIVERSE}

Inspired by the COBE results, a series of ground and balloon-based experiments measured cosmic microwave background anisotropies on smaller angular scales over the next decade. 
The primary goal of these second generation experiments was to measure the scale of the first acoustic peak,  which COBE did not have sufficient angular resolution to resolve. 
The first peak in the anisotropy was tentatively observed by the Toco experiment and the result was confirmed by the MAXIMA and BOOMERanG experiments.\cite{ref:flat}. 
These measurements demonstrated that the Universe is approximately flat and were able to rule out cosmic strings or textures as a major component of cosmic structure formation, and
thus suggested cosmic inflation was the right theory of structure formation.
The second peak was tentatively detected by several experiments before being definitively detected by WMAP, which has also tentatively detected the third peak.
Further experiments such as ACBAR and CBI provided strong evidence for higher peaks and the
long anticipated Silk-damping tail.

\section{The Inflationary Universe}

Inflation remains a viable and compelling theoretical idea. 
There is almost perfect agreement between the cosmic microwave background (CMB) anisotropy
observations and theoretical predictions based
on inflation. 
This wonderful agreement has been interpreted by many, 
though not all, cosmologists as evidence that inflation actually happened. 
What would the bulk of cosmologists consider sufficient evidence?

A key prediction of inflationary models is the existence of a stochastic
background of gravitational waves or technically tensor modes.
These tensor modes or primordial gravitational waves are potentially detectable 
by their `$B$-mode' polarization signature in the CMB~\cite{ref:zald, ref:kam}. 
Observing the $B$-mode signature would provide smoking gun evidence to convince residual doubters and provide a window into the energy scale and perhaps the mechanism of inflation. 

Unfortunately, for the same reason that it is so useful - namely we have no solid idea of the energy scale of inflation,  it is possible that the tensor mode produced signal could be quite small amplitude making it difficult to detect them not only because of the sensitivity requirements but also essentially impossible because their polarized signal will be so buried in the foreground Galactic polarized emissions.

\begin{figure}
\centering
\includegraphics[width=1.0\textwidth]{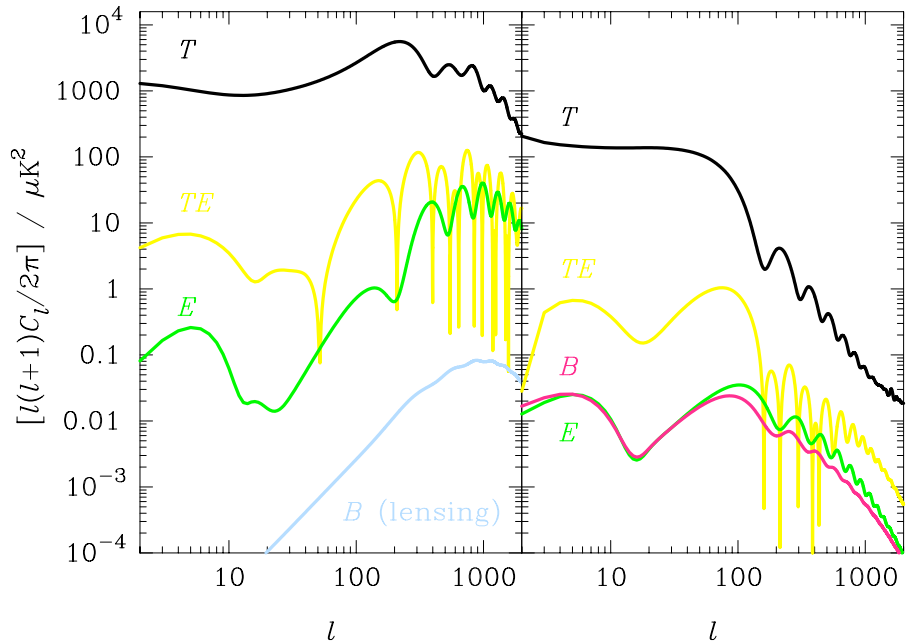}  
\vskip-0.5in   
\caption{Theoretical CMB angular power spectra.
On the right is the estimated effect from scalar density fluctuations:
T is the Temperature power spectrum.
TE is the Temperature cross E-mode Polarization power spectrum,
E is the E-mode Polarization power spectrum, and 
B is the resulting B-mode power spectrum from the gravitational lensing of the E-mode power spectrum from the scalar fluctuations. The levels are based upon the observed T power spectrum.
On the right are the maximum signals currently allowed for the tensor mode contributed power spectrum.
The signals from tensor modes could easily be orders of magnitude less.
Figure taken from article\cite{ref: Challinor}.}

\end{figure}

With COBE DMR observations of what appeared to be scale invariant fluctuations, 
I was convinced that Inflation was to be taken seriously and tested.
I got involved in figuring out how to move to the next level in 
probing and understanding Inflation as it seemed to be such a good description of the observed universe. 
I understood that Inflation should produce gravitational waves as well as scalar perturbations.
I had a mental picture of the process - Hawking or deSitter radiation and about a week after the COBE announcement I visited Cambridge at Hawking's invitation to give a lecture on COBE.
I spent time with Stephen trying to convince him to consider gravity waves from Inflation as a potential marker. 
Stephen was not interested. 
He was taken with his No-Boundary model which he declared produced  gravity waves at only a very low level.
This did not match my mental picture but I had not done the calculations.
Two days later I was teaching at a cosmology school in Amsterdam and staying in the same hotel with Paul Steinhardt. 
We would have breakfast together and walk a fair distance to school each day so I started discussing and arguing with Paul about how there should be significant (up to comparable with scalar modes) tensor modes and Paul saying that no the tensor modes were quite low and very much over shadowed by the scalar modes.
That is true for the limit of slow roll inflation which was Paul's invention and favorite.
But I kept after it and learned about the fluctuation mechanism from the quantum fluctuations of the inflaton field and convinced Paul that some models of inflation at very high energy could produce a significant gravity wave component.
In the end we wrote a paper on the general aspects\cite{ref:Smoot93}, a gravity prize essay, and a paper that included Mike Turner\cite{ref:Davis92} that first showed that there was a relationship between the fluctuations spectral index and the $r =$ tensor to scalar ratio.
This meant that there was a possibility both of testing inflation and determining its energy scale.

It soon became clear that looking for the polarization effects was the best way to separate the tensor and scalar modes.
I became quite excited about going out to observe the polarization.
However, as time has passed, I have been concerned that we may not 
be able to probe sufficiently deeply precisely because of the foreground issues combined 
with the large potential range for the inflation energy scale.
It is difficult not to draw parallels to the searches for cold dark matter
which so far have only probed the top of a very potentially large phase space.
I do think that we should continue experiments and probe and hope that the perturbation spectral tilt and predicted relationship to the tensor to scalar modes holds up and we are successful.

In the last fifteen years we have become used to improving high
precision information from CMB experiments. 
The temperature anisotropies have  provided vital information
 for  cosmology including  theories of the early Universe, models of structure
formation, primordial nucleosynthesis, neutrino masses, ages of the
oldest stars and so on. 

So my first thoughts were to continue to exploit the CMB as much as possible and ask 
what else could give us information on Inflationary models?
Clearly we have come to understand that there is information in:

(1) non-Gaussianity in the temperature anisotropies (which is surely there and will be exploited ). 
The simplest forms of inflationary models, e.g. slow-roll inflationary models, predict very small levels of non-Gaussianity.

(2) potential relic cosmic strings (or perhaps textures) from symmetry breaking at the end of inflation.
Many forms of inflation including brane-inflation do produce strings. 
These strings would leave the standard theory little changed as they would contribute little to structure formation and to the larger angular scale CMB power spectrum (current limit less than 10 per cent of the total signal). 
Here the level of probable existence and detectability leave something to be desired
as the low end of the energy scale of inflation is still an unknown and large in range.
The intrinsic scalar modes fluctuations provide a background floor for the CMB strings signature that is helped by going to smaller angular scales but this is also limited so that optical and gravity wave searches may ultimately provide the only hopes. 
Still the detection and study of a cosmic string (or superstring) would be revolutionary to our understanding.

\section{ Dark Energy and the Accelerating Universe}

The discovery of an accelerating Universe\cite{ref:perlmutter}, \cite{ref:riess} in 1998
had a high acceptance within a couple of years.
The $\Lambda$CDM model was quickly resurrected as an adequate description of cosmological observations.
Things might have stabilized there but Mike Turner coined the phrase `Dark Energy' to characterize this present day accelerating universe.
When I asked Mike about how he picked the phrase, Mike retorted,
``the first thing was to not call it the cosmological constant or vacuum energy or the answer would already be assumed. 
Then it came to what was the best short phrase."
Mike's kick to not assume
 has led to an explosion of papers on the
phenomenology of `dark energy' (for a recent review see {\it e.g.}
\cite{ref:copeland}). 

On the observational side a large number of
ambitious projects have been proposed to constrain the equation of
state of dark energy and its possible evolution (summarized concisely
in the {\it Report of the Dark Energy Task Force},
\cite{ref:albrecht}).

\section{Conclusions}

We have seen wonderful and exciting progress in the last fifteen years 
in our understanding of cosmological  models and determining cosmological parameters. 
We hope and nervously anticipate that events will continue to impress and delight us.

\subsection{Future CMB Experiments}
The big event and mission on the horizon is the Planck Mission scheduled for launch 
in about one year. 
We all hope that its observations will propel the field forward substantially.

Post Planck, most CMB experiments are targeting either high angular
resolution observations with focus on secondary anisotropies (the
Sunyaev-Zeldovich effect in particular) or low-noise (very sensitive) observations
of B-mode polarization anisotropies.  
Pending results from these,  a case may arise for a CMB polarization satellite mission.
The hints of non-Gaussianity now being discussed or the less likely detection of cosmic
strings could motivate a new generation of CMB or other observations that could provide breakthroughs in our understanding of cosmology.

The search for anisotropies in the cosmic microwave background 
began in 1964 with the discovery by Penzias and Wilson\cite{ref:Penzias65}
as noted by the quotation in their paper that the radiation was isotropic to the 10\%\ level.
In the succeeding forty years we have improved the measurements 
and observations by five orders of magnitude and come to discover and 
utilize the CMB anisotropies very effectively in understanding cosmology. 
We can congratulate our community on its remarkable progress.
The field has undergone a revolution in the last decade but there was a lot of ground work 
both experimental (observational) and theoretical that lead to that revolution.

\subsection{Understanding Dark Energy} 
This is a fundamentally important issue and one that is difficult to tackle observationally.
There will need to be more than one approach to tackle this and be sure the outcome is the correct result.
Understanding dark energy will require major projects and programs.
The CMB observations will provide the background cosmology for the effort.

\acknowledgments
This work was supported  by US DOE and the Berkeley Center for Cosmological Physics at the University of California at Berkeley. 
Grazie to Guido Chincarini for the valuable invitation and encouragement.
I am pleased that the volume is dedicated to Dennis Sciama who provided me with insight and encouragement.

\end{document}